# Hexagonal network of photocurrent enhancement in few-layer graphene/InGaN quantum dot junctions

*Guanghui Cheng*,*,[†,‡,§] *Zijing Jin*,[†,§] *Chunyu Zhao*,[⊥] *Chengjie Zhou*,[†] *Baikui Li*,[¶] and *Jiannong Wang*[*,†]

[†]Department of Physics, the Hong Kong University of Science and Technology, Clear Water Bay, Kowloon, Hong Kong

[‡]Department of Physics, University of Science and Technology of China, Hefei, Anhui 230026, China

[⊥]Department of Electronic and Computer Engineering, the Hong Kong University of Science and Technology, Clear Water Bay, Kowloon, Hong Kong

[¶]College of Physics and Optoelectronic Engineering, Shenzhen University, Nanhai Ave 3688, Shenzhen, China

[§]G.C. and Z. J. contributed equally to this work.

*Corresponding Authors. Emails: cghui@ustc.edu.cn; phjwang@ust.hk

**Abstract: Strain in two-dimensional (2D) materials has attracted particular attention owing to the remarkable modification of electronic and optical properties. However, emergent electromechanical phenomena and hidden mechanisms, such as strain-superlattice-induced topological states or flexoelectricity under strain gradient, remain under debate. Here, using scanning photocurrent microscopy, we observe significant photocurrent enhancement in**

**hybrid vertical junction devices made of strained few-layer graphene and InGaN quantum dots. Optoelectronic response and photoluminescence measurements demonstrate a possible mechanism closely tied to the flexoelectric effect in few-layer graphene, where the strain can induce a lateral built-in electric field and assist the separation of electron-hole pairs. Photocurrent mapping reveals an unprecedentedly ordered hexagonal network, suggesting the potential to create a superlattice by strain engineering. Our work provides insights into optoelectronic phenomena in the presence of strain and paves the way for practical applications associated with strained 2D materials.**

KEYWORDS: strain, graphene, InGaN quantum dots, optoelectronics, flexoelectricity, electromechanical coupling

Mechanical strain provides an extra degree of freedom to manipulate the intrinsic electronic and optical properties, as well as structural phase transitions[1-11]. Of particular interest are the two-dimensional (2D) materials, which can withstand deformation over 10 % before fracture, substantially larger than the bulk counterparts[1, 12]. The flexibility to create strain in 2D materials has stimulated much effort with numerous strain-induced phenomena. For example, modification of the bandgap[2, 3] and structural phase transitions[4] are found in 2D semiconductors in the presence of strain. In gapless graphene, strain can open an energy gap with pseudogauge fields and topological behaviors[6, 7, 9-11]. Mechanical deformation is thus an effective approach to control the physical properties and phase transitions of 2D materials.

The flexoelectric effect is one of the critical electromechanical couplings behind strain-related phenomena. Compared to the piezoelectricity under uniform strain, flexoelectricity describes the spontaneous electric polarization induced by the strain gradient. Recent advances in 2D materials and nanotechnology allow strain gradients to universally exist at corrugations[10, 13], domain walls[14] or dislocations[15], which are usually orders of magnitude stronger than macroscopic bending[16]. Such a large strain gradient at the nanoscale makes the flexoelectric effect increasingly dominant in the physical properties. Among 2D materials, nonpiezoelectric graphene is an ideal platform to study the flexoelectric effect[17-19]. The strain gradient can change the ionic positions and lead to asymmetric distributions of the electron density in graphene[17, 18], which can be exploited to control optoelectronic responses. Therefore, experimental manifestations are highly desired. On the other hand, traditional piezoelectricity measurements such as piezoelectric force microscopy are generally used to characterize the flexoelectric effect[14, 20, 21]. However, it is challenging to decouple the intrinsic flexoelectric response from the substrate effect or electrostatic interactions[14, 20]. In this regard, the exploration of optoelectronic phenomena in the presence of strain can shed light on the fundamental understanding of electromechanical couplings. In this work, we combine strained few-layer graphene (FLG) with InGaN quantum dots (QDs), a representative optoelectronic material, and investigate the strain-related optoelectronic behaviors. The InGaN QDs can serve as an opto-active exciton reservoir due to efficient photon-exciton conversion and long-lifetime of excitons. We find significant enhancement of the photocurrent which arises from the strain-gradient-induced polarization and lateral built-in field in FLG. The demonstrated flexoelectricity in FLG serves as a new photocurrent generation mechanism, in additional to the reported photovoltaic[22], thermoelectric[23, 24] and bolometric mechanisms[25] in graphene devices. The photocurrent mapping reveals an unprecedentedly ordered hexagonal network, suggesting the potential of constructing long-range superstructures by modulating the spatial distribution of local strain in 2D materials.

Hybrid structures are prepared using dry-transfer techniques to integrate exfoliated FLG and MOCVD-grown InGaN QDs on sapphire. The structural characterizations of InGaN QDs are shown in Figure S1 in the Supporting Information. The deposition of an ~40 nm amorphous $Si_3N_4$ film is performed by PECVD. Due to different thermal expansion coefficients, strain can be readily introduced to FLG by thermal cycles during deposition. (see Methods for details)

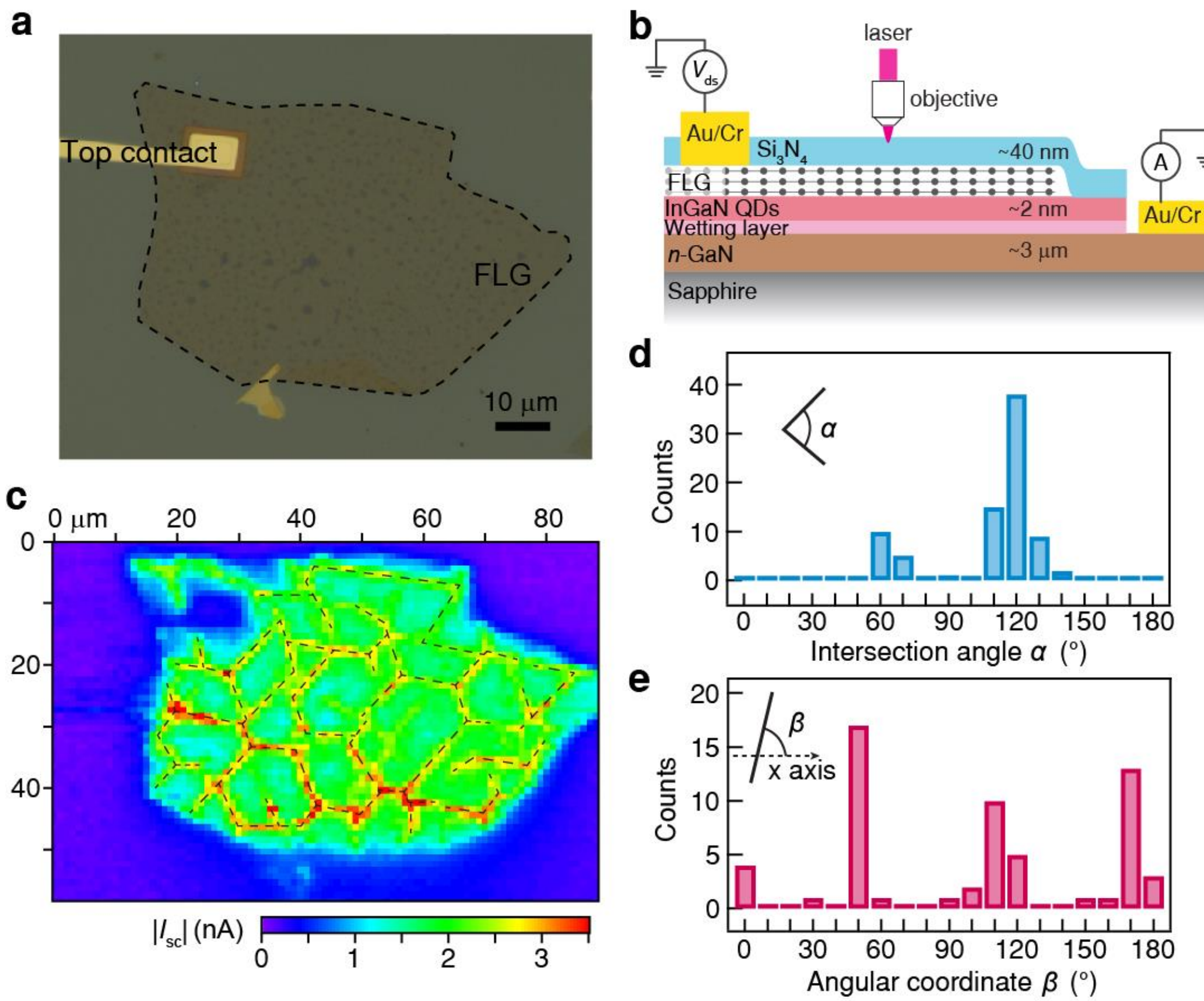

Figure 1. Scanning photocurrent microscopy of the FLG/InGaN QD junction. (a) Optical micrograph of FLG/InGaN QD junction device 1 with a transparent $Si_3N_4$ layer on top. A black dashed line outlines the FLG. The substrate is composed of InGaN QDs/wetting layer/*n*-GaN/sapphire. (b) Schematic illustration of the cross-section of the vertical junction device based on FLG/InGaN QDs. A 400-nm laser is focused onto the sample surface with a 100× objective. (c) Mapping of the short-circuit photocurrent $I_{sc}$ of the same area as in panel (a). Dashed lines mark the network of enhanced photocurrent. Laser power of 30 μW is used. (d,e) Histograms of the distributions of the intersection angle $\alpha$ and the line angular coordinate $\beta$ for the patterns in panel (c). Only $\alpha$ and $\beta$ in the range of 0° ~ 180° are included here. $\alpha$ shows primary distributions at 60° and 120°, while $\beta$ shows a periodicity of ~60°.

Figure 1a shows the optical micrograph of a typical vertical junction device based on FLG/InGaN QDs (device 1). The thickness of FLG is ~6.4 nm, as confirmed by atomic force microscopy (AFM). Two-terminal electrodes are fabricated with a bottom ohmic contact to *n*-GaN (Figure S2 in the Supporting Information) and a top contact to FLG, with the latter shown by the yellow bar in the image. The schematic device cross-section and the photocurrent mapping setup are illustrated in Figure 1b. A transparent $Si_3N_4$ layer is deposited to isolate the top electrode from the substrate and particularly to induce strain in FLG. Due to the negative thermal expansion coefficient of FLG compared to the positive value of $Si_3N_4$[26, 27], it is expected that a compressive strain is induced in FLG during cooling after the growth of the $Si_3N_4$ film at 300 °C. We employ scanning photocurrent microscopy to probe the spatially resolved photocurrent response of the junction (see Methods). A 400-nm laser is focused onto the sample surface with a 100× objective. Samples are placed in an optical cryostat, and the data are obtained at a temperature of 6 K unless otherwise specified.

3.1-eV photons transmit through the junction and are absorbed by InGaN QDs (optical bandgap ~ 2.38 eV estimated from the absorption and photoluminescence spectra in Figure S3 and S4 in the Supporting Information). The asymmetric potential across the vertical FLG/InGaN QD junction leads to the separation of photoexcited electron-hole pairs and a finite short-circuit photocurrent $I_{sc}$. Typical spatial mapping of $I_{sc}$ under a laser power of 30 μW is shown in Figure 1c. Remarkably, the spatially resolved $I_{sc}$ shows a profound enhancement at particular positions, marked by the dashed lines The filtered mapping of $I_{sc}$ with a threshold value of 1.9 nA is shown in Figure S5 in the Supporting Information. $I_{sc}$ along the lines (red/yellow colors) can be up to four times larger than the interior parts (cyan/green colors). Considering the limited spatial resolution of the laser spot, this enhancement is underestimated. More strikingly, these line-like enhanced $I_{sc}$ regions form a hexagonal network structure with a length scale of ~10 μm.

An intuitive statistical analysis of the intersection angle $\alpha$ and line angular coordinate $\beta$ is performed, as shown in Figures 1d and 1e. The angle $\alpha$ is distributed primarily at ~60° and ~120°, indicating an essential hexagonal structure. Consistent results are observed for $\beta$, showing 60° angular periodicity. The hexagonal order is further verified by Fourier transform on the photocurrent image shown in Figure S5 in the Supporting Information. Such a hexagonal pattern is also observed in the photocurrent mapping of another FLG/InGaN QD junction with the same structure (Figure S6 in the Supporting Information). These observations suggest a sixfold rotational symmetry of the pattern, which suggests the potential to form a hexagonal superlattice. Such a pattern excludes the origin of domain boundaries in graphene, which are reported to break translational or rotational symmetry[28]. We further perform photoluminescence (PL) mapping on device 1 (Figure S4 in the Supporting Information) and demonstrate that the structural stoichiometric nonuniformity or local defects[29] in InGaN QDs do not play an important role in the observed photocurrent enhancement pattern.

The hexagonal pattern is consistent with the graphene honeycomb lattice, which suggests its origin of strain engineering in FLG. Out-of-plane deformation (e.g., wrinkles) is usually reported in strained graphene by the interlayer sliding[9-11, 13, 30-32]. The distribution of strain depends on the detailed shape and size of the deformation[32]. We thus perform AFM measurements on device 1 (Figure S7 in the Supporting Information). Indeed, the observed wrinkles match up with the observed photocurrent pattern, supporting the strain-related origin. Note that some areas show enhanced photocurrent but no distinguishable wrinkles in the AFM images, possibly due to the in-plane strain nature. The comparison between topography and photocurrent pattern suggests that both in-plane and out-of-plane strains contribute to the photocurrent enhancement observed. Such photocurrent enhancement phenomena along wrinkles are also observed in other two samples (Figure S6 in the Supporting Information). Not only at wrinkles, we also observed photocurrent enhancement at the bubble region where significant strain is present (Figure S6 in the Supporting Information).

Furthermore, temperature dependent photocurrent shows a hysteresis loop upon warming and cooling the device (Figure S8 in the Supporting Information). Such hysteresis behavior corresponds to the typical feature of strain[33, 34], suggesting the critical role of strain in the photocurrent enhancement.

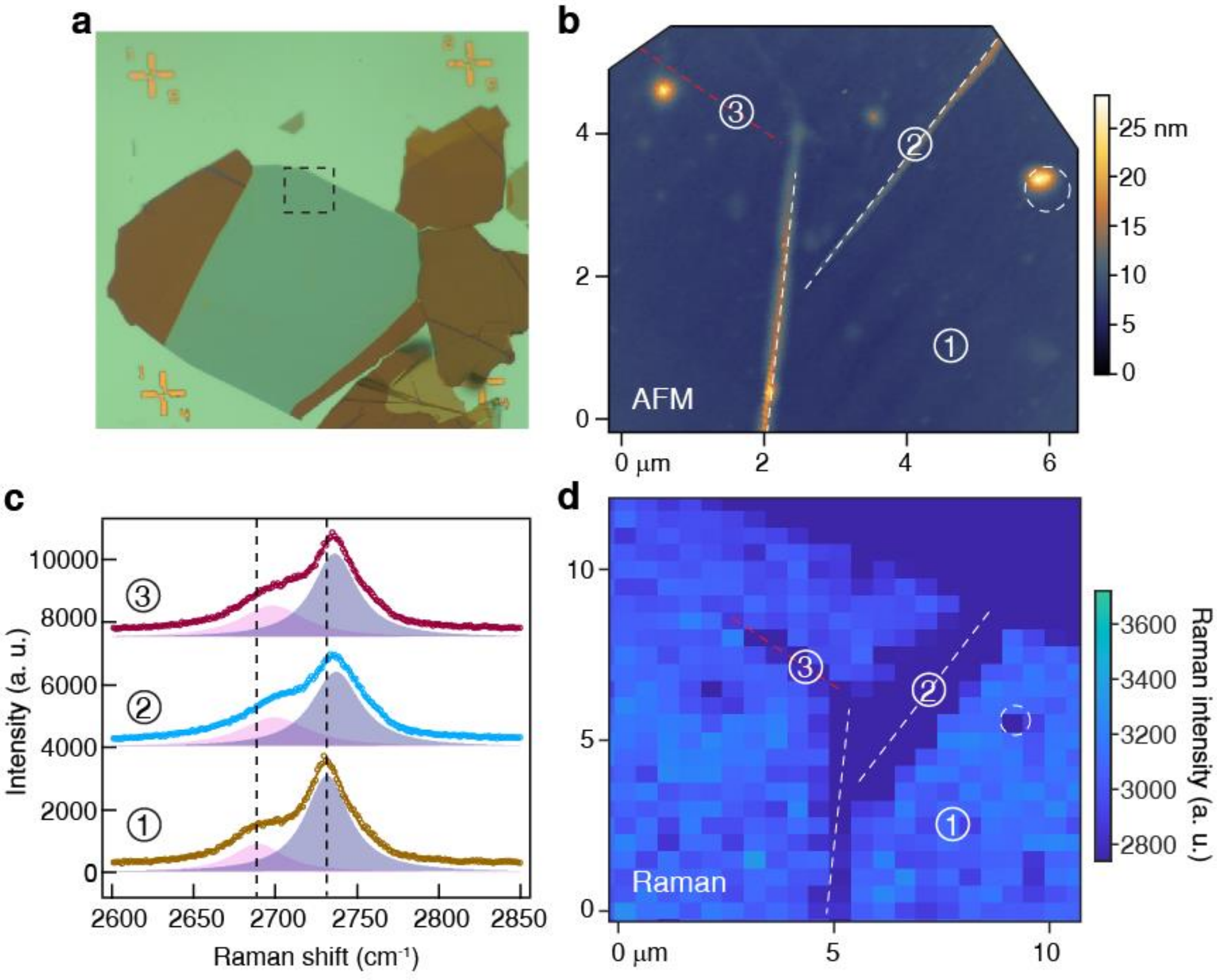

Figure 2. Characterization of the strains in the FLG/InGaN QD junction. (a) Optical micrograph of FLG/InGaN QD device 2. The visible flakes are FLG. The substrate is composed of InGaN QDs/wetting layer/*n*-GaN/sapphire. (b) AFM image of a selected area denoted by a dashed rectangle in panel (a). Red

and white dashed lines mark the positions with strain. (c) Raman spectra of the 2D mode of FLG measured at the selected positions ①, ②, ③ denoted by the circled numbers in panel (b). Original data are plotted as circles and fitted by two primary Lorentzian components indicated by the shaded peaks. (d) Scanning image of Raman intensity for the 2D mode of FLG. Red and white dashed lines mark the same strained positions as in panel (b).

To characterize the strain in the hybrid structure, Raman spectroscopy has been measured where the strain leads to peak shifts and intensity modulation of typical graphene Raman modes[30,31]. Figure 2a shows an optical micrograph of FLG/InGaN QD device 2 with the same fabrication processes as device 1. The AFM measurements performed in the dashed rectangular area are shown in Figure 2b. White and red dashed lines represent the wrinkles and undistinguishable in-plane strain, respectively. Both types of strain strongly affect the 2D Raman mode of FLG, as shown in Figure 2c. Compared to the unstrained position ①, we observed blueshifts for both the primary peak (5.5 cm$^{-1}$ and 4.7 cm$^{-1}$) and shoulder peak (10.1 cm$^{-1}$, 8.9 cm$^{-1}$) for positions ② and ③, respectively. Based on the reported dependence of Raman spectra on the strain[35], the strain in our structure is determined to be the compressive type with estimated strengths of ~0.06 % and ~0.05 % for positions ② and ③, respectively. Moreover, the spatial distribution of strain can be visualized in Raman intensity mapping of the 2D mode of FLG (Figure 2d). Red and white dashed lines mark the same strained positions as in the AFM image in Figure 2b. The above characterizations indicate that compressive strain is induced at certain locations in the FLG of our hybrid structure. Further experiments on devices 3 and 4 show consistent results between the AFM, Raman and photocurrent images, suggesting a strain-related photocurrent enhancement (Figure S6 in the Supporting Information).

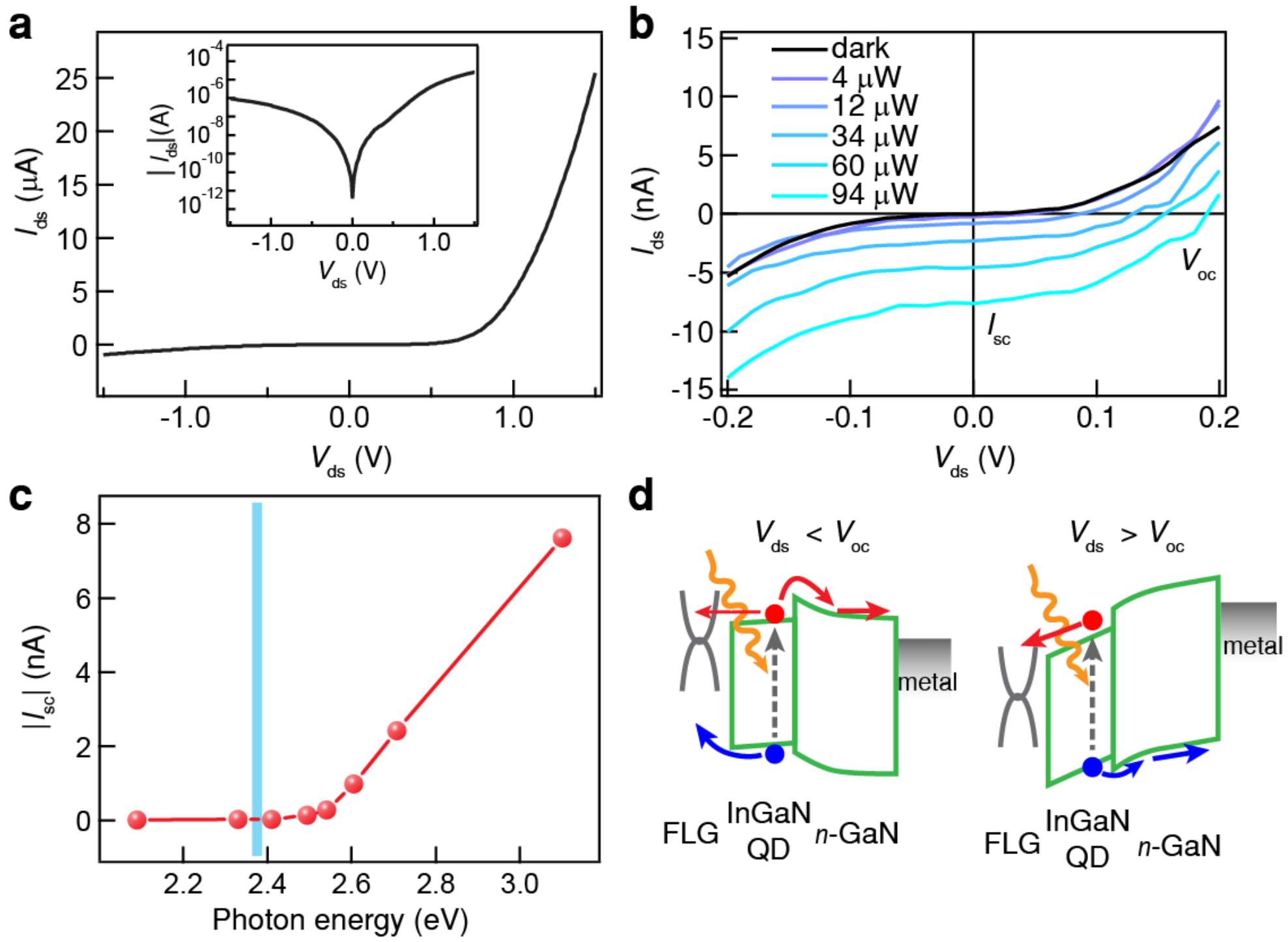

Figure 3. Optoelectronic characterization of FLG/InGaN QD junction device 1. (a) Current-voltage characteristics of the FLG/InGaN QD junction. The inset shows the absolute current-voltage curve in semi-logarithmic scale. (b) Current-voltage curves under a 400-nm laser with the total illumination power indicated. The short-circuit photocurrent $I_{sc}$ and open-circuit photovoltaic $V_{oc}$ can be extracted. (c) $I_{sc}$ normalized by the incident photon numbers as a function of photon energy. The resonant energy of InGaN QDs is denoted by the cyan line. (d) Schematic band diagrams and photocurrent generation processes for the cases of $V_{ds} < V_{oc}$ and $V_{ds} > V_{oc}$. Electrons and holes are denoted by red and blue balls, respectively.

To gain an in-depth understanding of the enhanced photocurrent at strain locations, we explore the photocurrent generation mechanism in the FLG/InGaN QD junction device. Two-terminal *I-V* characteristics are obtained with the bottom contact grounded, as shown in Figure 3a. Considerable current rectification is observed. Figure 3b shows the photoresponse of *I-V* curves under a 400-nm laser with the total illumination power indicated. The laser spot is fixed at the center of the FLG flake, covering both unstrained and strained areas due to the spatial distribution of the laser spot. As it can be seen, short-circuit photocurrent $I_{sc}$ is up to ~ -8 nA, and open-circuit photovoltaic $V_{oc}$ is up to ~ 0.2 V as the illumination power increases. Figure 3c shows $I_{sc}$ as a function of excitation photon energies normalized by the incident photon numbers. The monotonic drop of $I_{sc}$ with decreasing photon energy agrees with the interband transition of InGaN QDs (optical bandgap ~2.38 eV). $I_{sc}$ reaches zero at the resonant energy of InGaN QDs because of the presence of energy barriers for the band-edge photocarriers injection to FLG and GaN (see detailed analysis in Supporting Information Note 1). Such negligible $I_{sc}$ at lower photon energies also suggests that the photogenerated charges in graphene can hardly contribute to the photocurrent due to its limited absorption[36] and ultrashort lifetimes of femtoseconds to picoseconds[37]. Further controlled experiment on a strained FLG device without InGaN QDs (Figure S9 in the Supporting Information) suggests the indispensable role of InGaN QDs in photocurrent generation in this work. The band diagrams are depicted in Figure 3d, based on the reported affinity energies, bandgaps and InGaN polarization field[38, 39] (see detailed analysis in Supporting Information Note 2). The thin InGaN wetting layer is not considered here due to the fast relaxation of photocarriers to the lower-lying QD levels[40]. For $V_{ds} < V_{oc}$ including the short-circuit case, the negative polarity of the photocurrent indicates that the photoholes and photoelectrons generated in the InGaN QDs are collected by the top FLG and the bottom metal electrode, respectively. A portion of the photoelectrons can be injected into the FLG and recombine with holes. For $V_{ds} > V_{oc}$, the external electric field strongly tilts the bands. Photoexcited electron-hole pairs in InGaN QDs separate following the external field, leading to a positive photocurrent.

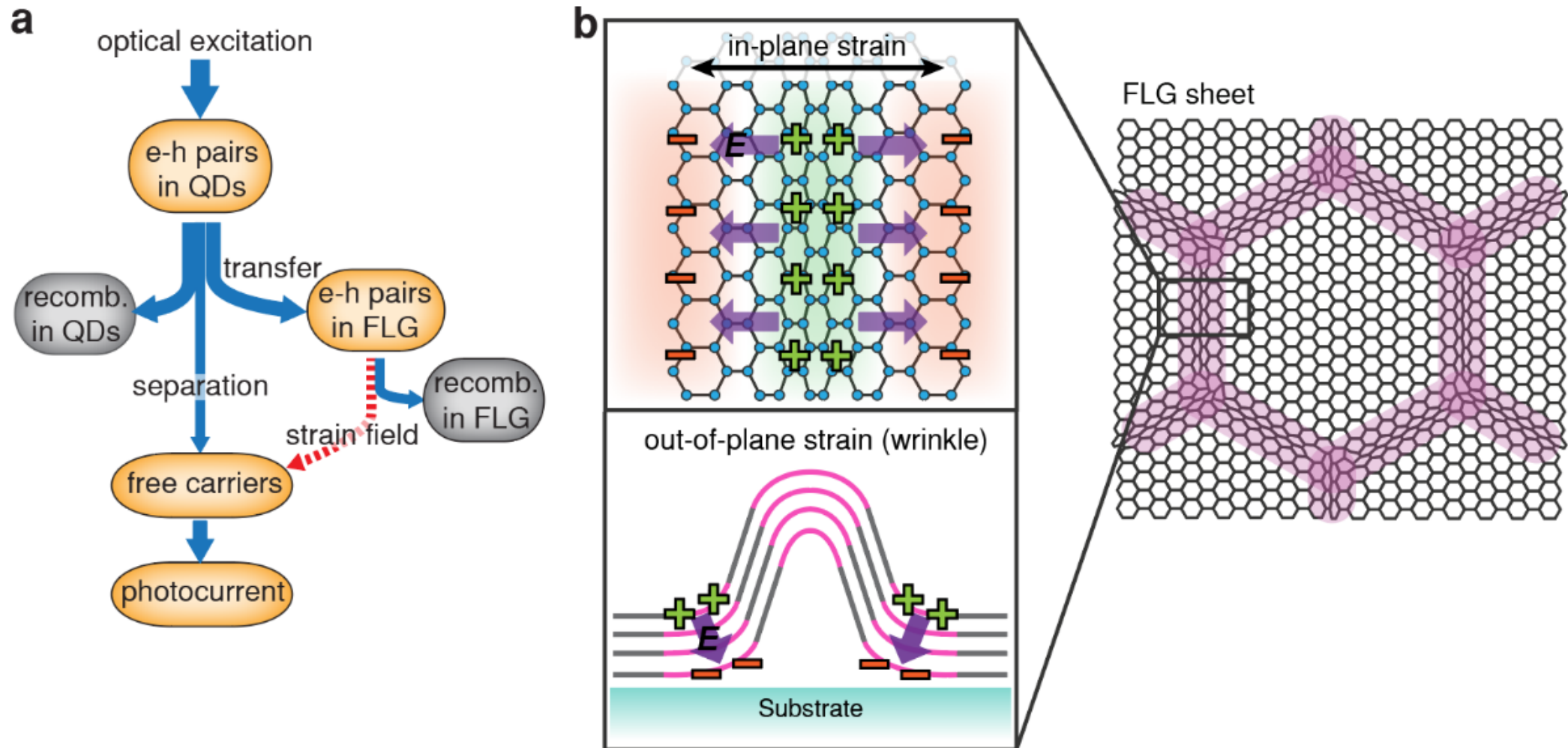

Figure 4. (a) Flow diagram of the FLG/InGaN QD junction summarizing the pathways to generate photocurrent. The red dashed arrow denotes the electron-hole separation under the strain-induced field. (b) Schematics of the flexoelectric polarization induced by in-plane and out-of-plane strains in FLG. Asymmetric distributions of positive (plus symbols) and negative (minus symbols) charges are produced along the strain gradient, forming a depolarization field (purple arrows) opposite to the flexoelectric polarization. This built-in field assists the separation of electron-hole pairs and enhances the photocurrent. The field direction is adapted from Refs. 41 and 42. The strain accumulates along the stiff direction of the FLG lattice and forms a hexagonal network.

Figure 4a shows the flow diagram of photocarrier conversion processes in the FLG/InGaN QD junction. Following blue arrows, the optically excited electron-hole pairs (excitons) in InGaN QDs are partially separated and contribute to the photocurrent. Other pairs go through radiative or nonradiative recombination processes either in QDs or in FLG. The exciton transfer from QDs to FLG and then recombination in FLG are verified by the decreased radiative lifetime in the FLG/InGaN QD junction using time-resolved PL measurements (Figure S10 in the Supporting Information).

Based on the above characterization of strain and optoelectronic responses, the enhancement of photocurrent can now be understood based on the flexoelectricity under in-plane and out-of-plane strains, as shown in Figure 4b. The inhomogeneous interactions of FLG with the $Si_3N_4$ film cause strain to accumulate along the stiff directions of the FLG lattice, presumably along the armchair or zigzag directions with a lower formation energy[43-45], in agreement with previous experimental observations[31]. This is also consistent with the observed hexagonal network of the enhanced photocurrent. The structural deformation in the FLG lattice can change the local charge distributions[9, 17, 18, 46] and generate an electric field along the in-plane or out-of-plane strain gradient. This strain-induced built-in field can efficiently separate electron-hole pairs transferred from QDs to FLG, which is in contrast to regions without strain where the transferred electron-hole pairs tend to recombine nonradiatively in FLG[47]. The separated carriers can contribute to the photocurrent under the asymmetric potential of the vertical junction, resulting in a hexagonal network of photocurrent enhancement (see red dashed arrow in Figure 4a).

The strain gradient can alter the local electronic properties and generate a photocurrent via other mechanisms, such as photothermoelectric or bolometric effect. However, the photothermoelectric effect is sensitive to the sub-bandgap illumination[48], while in our system the photocurrent diminishes at photon energies lower than the resonant energy of InGaN QDs (see Fig. 3c). Moreover, the photothermoelectric-related photocurrent can be generally observed close to the device contact where a junction is formed between two materials with different Seeback coefficients[48, 49]. However, such an effect is not observed in our device as shown Fig. 1c, where the photocurrent gradually decreases when getting close to the contact. The bolometric effect is basically a photo-conductance effect which plays a role only in biased devices[50] while in our case we measure short-circuit photocurrent at FLG/InGaN QD junction. On the other hand, the flexoelectric photocurrent becomes increasingly prominent as the system size diminishes, especially at wrinkles of 2D materials. As a result, it can be regarded as a novel photovoltaic mechanism, where the local electric field is produced by the strain-gradient-induced built-in field.

We can now revisit the photocurrent mapping in Fig. 1c, where the flexoelectric photocurrent dominates the strained areas, and its tail effect together with other possible mechanisms discussed above contribute to the photocurrent between the strain areas.

The photocurrent generation in terms of the flexoelectric effect in the FLG/InGaN QD junction can be estimated as[51]

$$I_{\mathrm{sc}} = (1+\beta) R I_{\mathrm{light}},$$

where $\beta$ is the flexoelectric-photocurrent coefficient ($\beta = 0$ refers to a situation without flexoelectric effect), $R$ is the conventional optoelectronic responsivity (including the light absorption), $I_{\mathrm{light}}$ is the light intensity. $\beta$ is proportional to the flexoelectric polarization $P_{\mathrm{flexo}}$, which can be expressed as[52]

$$\beta \propto P_{\mathrm{flexo}} = \mu\eta,$$

where $\mu$ is the flexoelectric coefficient and $\eta$ is the strain gradient of the strained FLG. In our FLG/InGaN QD junction, we obtained $\beta \sim 4.6$ for the strained FLG region (see detailed calculation in Supporting Information Note 3). Note that the overall flexoelectric-photocurrent strength depends not only on the strain gradient ($\eta$), but also on the flexoelectric coefficient ($\mu$) of the material and the conventional optoelectronic responsivity ($R$) of the junction.

The piezoelectricity can be induced in graphene by breaking its centrosymmetry under certain conditions including structural nanopatterning[53], selective surface adsorption of atoms (doping)[54], interaction with the substrate[20], or local structural deformation (e.g., defects)[55]. Except nanopatterning, other effects may exist in our devices. However, they can only play minor roles since their random distribution nature is contradictory with the observed ordered hexagonal pattern of the photocurrent enhancement. To fully understand the role of piezoelectric effect, future experiments are needed, such as piezoelectric force microscopy and inline electron holography.

We note that the observed hexagonal pattern suggests the potential to realize a strain superlattice in a more controlled manner[56]. Similar to moiré physics[57, 58], the periodic potential enabled by the strain superlattice can be a robust and feasible way to discover and control new phases of matter in materials research[6, 9, 59].

In conclusion, we demonstrate significant strain-induced photocurrent enhancement in the FLG/InGaN QD junction, which is most probably associated with a flexoelectric-driven mechanism. Our findings provide insights into the optoelectronic behaviors of 2D materials in the presence of a strain gradient. The unraveled electromechanical coupling provides exciting implications for electromechanically active materials and nanodevices based on graphene.

## Methods

*InGaN QD growth.* The InGaN QD samples were grown in an Aixtron close-coupled showerhead (CCS) MOCVD reactor. Trimethylgallium (TMGa), trimethylindium (TMIn), and ammonia ($NH_3$) were employed as sources for Ga, In, and elemental N, respectively. Thermal cleaning was performed to desorb the residual native oxides at the surface of the c-plane sapphire substrate followed by a GaN nucleation layer, an unintentionally doped GaN layer, and a 3 μm silicon-doped GaN layer to form a GaN pseudosubstrate. Prior to the growth of InGaN QDs, a 600 nm silicon-doped GaN connecting layer was grown to bury the regrowth interface and provide a smooth surface. An uncapped layer of InGaN QDs was grown at 670 °C with a V-to-III ratio of $1.02 \times 10^4$. The self-assembled quantum dots are formed by the Stranski−Krastanov (SK) growth mode. The structural characterizations of InGaN QDs can be found in our previous work[38] and in Figure S1 in the Supporting Information.

*Vertical junction fabrication.* Before transferring FLG, we first used optical lithography and $BCl_3/Cl_2$-based ICP etching to remove the quantum dot layer and wetting layer to expose the *n*-GaN layer near the sample edge. The bottom contact is fabricated by electron-beam lithography and metal deposition to make contact with *n*-GaN, followed by the surface cleaning process. FLG flakes are mechanically exfoliated onto a thin layer of polydimethylsiloxane (PDMS), and the thicknesses are confirmed by optical contrast and later by AFM. With the aid of an optical microscope and micromanipulators, the selected FLG flake was placed in contact with the QDs to form the hybrid structure depicted in the main text. Transfer processes were carried out in a nitrogen-filled glovebox. A rectangle-shaped top electrode is fabricated by standard electron-beam lithography to make contact with FLG and later acts as the anti-etching pad. A 40-nm amorphous $Si_3N_4$ film was then deposited on the whole sample surface by PECVD with a growth temperature of 300 °C. Finally, we used electron-beam lithography, reactive ion etching, and metal deposition to etch through $Si_3N_4$ and make contact with the top FLG, acting as the top contact.

*Optical/optoelectronic characterization and scanning photocurrent microscopy.* Samples are placed in an optical cryostat and the data are obtained at a temperature of 6 K unless otherwise specified. A 400-nm laser was focused onto the sample surface with a 100× objective. For PL measurements, emitting photons are detected by monochromator and photomultiplier tubes. For the time-resolved PL measurement, a pulsed 400-nm femtosecond laser and streak camera system are employed for PL excitation and detection, respectively. The scanning photocurrent and *I-V* characteristics are measured by the AC lock-in technique at 17 Hz and a Keithley 2636B source meter, respectively. The two methods show consistent results. Precise motorized actuators are employed to mechanically move the sample cryostat to realize the scanning function. The spatial resolution of the scanning photocurrent system is determined by the diameter of the focused laser spot ~0.8 μm[60].

ASSOCIATED CONTENT

**Supporting Information**

Dependence of photocurrent on the excitation energy; Band diagram analysis of the hybrid FLG/InGaN QDs system; Estimation of photocurrent mechanism in terms of the flexoelectric effect; Structural characterizations of InGaN self-assembled quantum dots; Current-voltage characteristics of the bottom contact with n-GaN; Band diagrams and photocurrent generation processes; PL characterization of the FLG/InGaN QD junction; Analysis of the hexagonal feature; AFM characterization of the wrinkles; AFM and photocurrent results in other two samples; Temperature dependence of the photocurrent; Controlled experiments without InGaN QDs; Decay curves of the PL intensity.

AUTHOR INFORMATION

**Corresponding Authors**

*(G.H.C.) Email: cghui@ustc.edu.cn; (J.N.W.) Email: phjwang@ust.hk

**Notes**

The authors declare no competing financial interests.

ACKNOWLEDGMENTS

We thank Prof Kei May Lau of the Department of Electronic and Computer Engineering of HKUST for supporting InGaN QD sample growth. This work is supported by the Research Grants Council of the Hong Kong SAR under grant nos. 16307019, C7036-17W-1, and AoE/P-701/20, and also in part by National Natural Science Foundation of China under Grant 62074103. We acknowledge the technical support from the HKUST Nano Fabrication Facility and Materials Characterization and Preparation Facility.

For Table of Contents Only

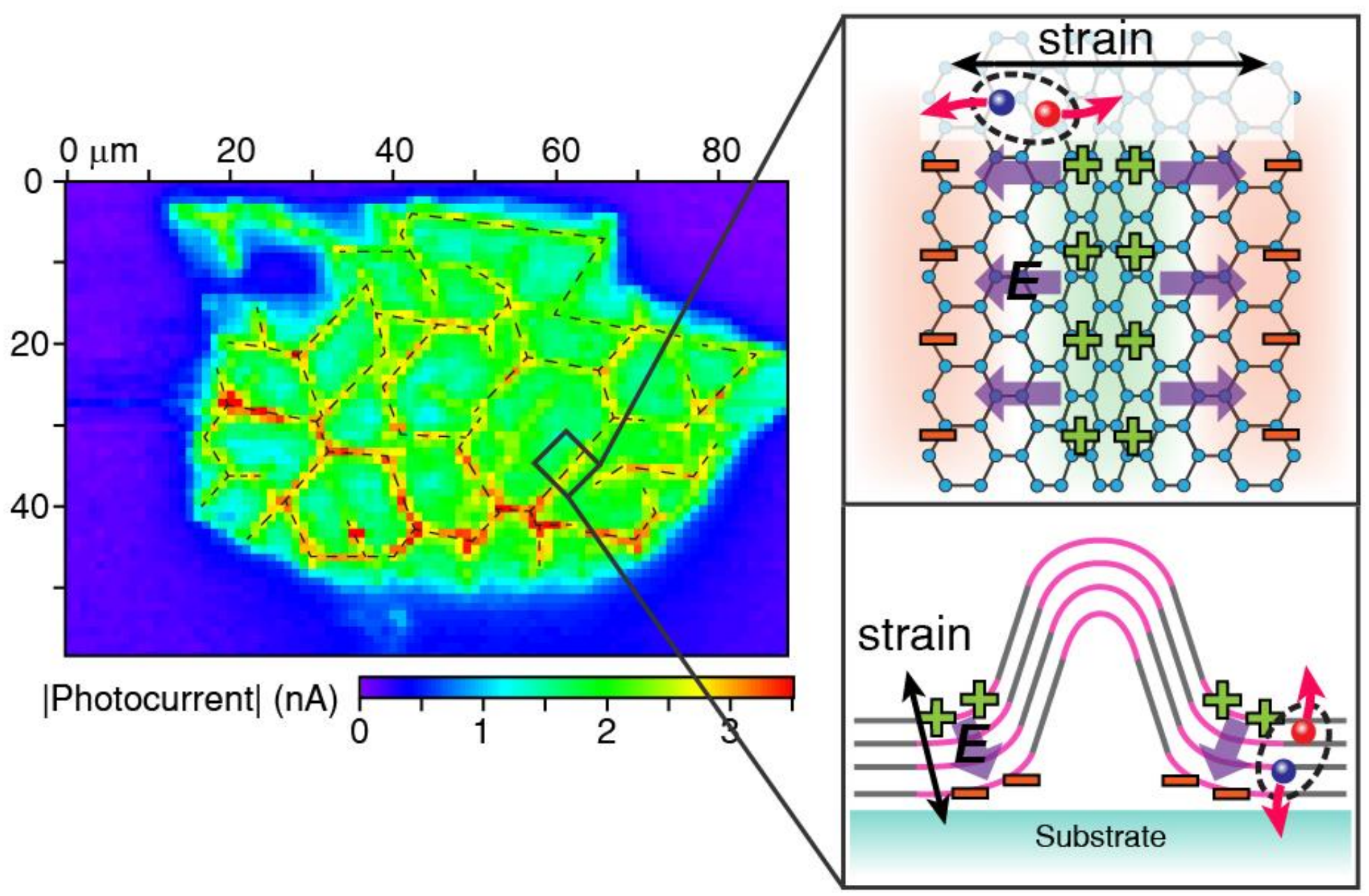